\documentstyle{article}
\begin{document}
\title{Density of states and magnetic susceptibilities on the
 octagonal tiling}
\author{A. Jagannathan}
\date{
Laboratoire de Physique des Solides
\cite{assoc},
 Universit\'{e} Paris-Sud,
91405 Orsay,
France }
\maketitle
\begin{abstract}
We study electronic properties as a function of the six types of local
environments found in the octagonal tiling. The density of states has
six characteristic forms, although the detailed structure differs from
site to site since no two sites are equivalent in a quasiperiodic
tiling. We present the site-dependent magnetic susceptibility of
electrons on this tiling, which also has six characteristic dependences.
We show the existence of a non-local spin susceptibility, which decays
with the square of the distance between sites and is the quasiperiodic
version of Ruderman-Kittel oscillations. These results are
obtained for a tight-binding Hamiltonian with pure
hopping.Finally, we investigate the
formation of local magnetic moments when electron-electron interactions
are included.
\end{abstract}

\vskip 1.5cm
This study of a tight-binding model on the two dimensional quasiperiodic
tiling with eight-fold symmetry, the octagonal tiling,
is aimed at understanding electronic properties in real space. There is
not much known about quasiperiodic Hamiltonians in two and higher
dimensions.
 To date, studies on quasiperiodic tight-binding models
have focused on the form of the density of states, on the response
of energy levels to changes of boundary condition (which gives a measure
of the degree of localization of the wavefunctions), and on the conductance
of such tilings (reviewed in \cite{fuj}). These studies
are important for understanding experiments on transport phenomena, and
for resolving whether quasicrystals are intrinsically metals or insulators.
However the properties studied are site-averaged properties, and are
characteristic of the tiling as a whole. Quasicrystals are interesting for
their $local$ properties as well, since they allow local atomic environments
which are different from those found in $periodic$ ordered structures.
One indication of site-dependent effects is given by nuclear magnetic
resonance and M\"ossbauer experiments \cite{hip} which find
that in magnetic quasicrystals, the value of the local magnetic
moment on a site appears to vary with the type of site.
There are no theoretical calculations to date that predict the distribution
of magnetic moments of ions on quasiperiodic structures. This motivates
our present study of
 local, site-dependent electronic properties in quasicrystals.

We begin with calculating the
 local density of states, within a non-interacting tight-binding model,
for each of the six local environments on the
octagonal tiling. Next we present the onsite magnetic susceptibility of
conduction electrons on this tiling. This is proportional to the spin
polarization density at the site in question when an external magnetic field
is applied. Another interesting quantity calculated
 is the nonlocal susceptibility, which tells how two magnetic moments
interact with each other as a function of their
positions within the quasicrystal. This changes sign as a function of the
distance, as
in a crystalline medium, the quasicrystalline version of Ruderman-Kittel
oscillations. Finally, we present results on the magnetic instability
induced by electron-electron interactions. Upon adding an onsite repulsion U
local moments appear on certain sites depending on the type of
site and the position of the Fermi level.

This study hopes as well to add to the scant intuition existing about
physical properties of quasiperiodic systems, which lack models, even
phenomenological, of their own. It is not the intent here to produce a
realistic band structure for a quasicrystal -- for which certainly far more
complex calculations using realistic parameters for atomic orbitals in
 binary and ternary alloys have been carried out \cite{may}.
Instead, we will consider possibly the simplest of models, whose
properties
we hope will not be overly model-specific. The octagonal
tiling (which is infrequently encountered in real materials) is appealing in
that it is the two-dimensional cousin of the 3d Penrose tilings used to
model a large class of quasicrystals, and is much closer mathematically
to these than is the 2d Penrose
tiling. Thus we hope that much of the discussion that follows will be
carried over to three dimensions.

The approximants of the octagonal tiling are generated by the projection
method described in \cite{dun}, which yields square pieces of the
infinite tiling, and allows for their periodic repetition in the x-y
plane. These approximants are obtained by taking successive rational
approximants of $\lambda={\sqrt 2}$, setting it to equal $O_k/O_{k+1}$
where
${1,2,5,12...O_k..}$ is the so-called Octonacci sequence \cite{dun}.
The number of sites in the square approximants depend on these Octonacci
numbers as well, and is 41, 239, 1393..etc for k=1,2,3...

The tightbinding model chosen in this work takes a single hopping
matrix element t=1 between sites connected by a bond in Fig.1. The six types
of site have been labeled in the figure.
This Hamiltonian possesses a symmetry $t \rightarrow -t$ (since the structure
shown is bipartite) and the energy band is thus symmetric about zero.
This is no longer the case when hopping is allowed across the short
diagonals of the rhombus, for instance, or as in the study of
vibrational spectra on these lattices, when diagonal terms appear
the Hamiltonian.

  We obtain first the local density of states (LDOS), using a continued
fraction expansion and recursion method \cite{hay}
for the Green's functions. The continued fraction is summed to infinite
order, in an approximation
used with good results in crystals and amorphous solids (reviewed in
\cite{span}) and seen to yield satisfactory
results in our present case.
 The form of the LDOS depends on the
site coordination number for its overall shape. Superposed on this is a
"fine structure" which varies from site to site, and shows
the intrinsically multifractal scaling property that is associated with
the spectra of quasiperiodic Hamiltonians, whether for the 1d
Fibonacci chain \cite{koh} or the vibrational spectrum on the octagonal
tiling \cite{los}. The present work is not concerned
so much with the fractal properties of the spectrum \cite{us} and of the
associated
spiky behavior of the susceptibilities, but rather with extracting
systematic variations obtained, statistically speaking, for sites of a
$given$ local environment.

The continued fraction representation of the Green's function
$G_{ii}(z)$ corresponding to a given site $i$ of the approximant can
 be written
\begin{equation} \label{ctfr}
G_{ii}(z) = \frac{1}{ z - \frac{b_1}{ z - \frac{b_2}{z - ...}}}
\end{equation}
where the parameters ${b_n}$ are the offdiagonal matrix elements in
the tridiagonalized form of the operator $(z-H)^{-1}$ (the
diagonal elements are zero due to the symmetric form of the energy spectrum
of the Hamiltonian). To calculate
the $b_n$ corresponding to a given site, we have shifted that site to the
center of the square tiling, and by repeatedly applying the Hamiltonian
to the original basis vector consisting of that site,
generated successive basis vectors in the tridiagonal basis,along with
the sequence of parameters
$b_n$. This procedure
yields a finite number of the parameters in the continued fraction in
Eq.\ref{ctfr}, as the procedure is stopped upon reaching the boundary.
The first twenty $b_n$ have been calculated for the 1393 site tiling in
this manner. One may truncate the continued fraction and evaluate
the density of states. A better approach is possible in view of the fact
that there are no gaps in the spectrum \cite{sire}. In this situation
it is easy
to show that if one replaces the exact $b_n$ by a fixed value $b_{\infty}$
the relation $b_{\infty} = \Delta^2/16$ holds (where $\Delta=2E_{max}$
is the band width). The band width can be inferred from $b_\infty$ and
vice versa, but of course neither is known $a priori$ \cite{note}. In
practice,
one can estimate $b_\infty$ by calculating the local density of states
$\rho_i(E) =-lim_{\delta \rightarrow 0} Im G_{ii}(E+i \delta)/\pi$
by summing the infinite continued fraction and requiring that $\rho$
be normalized to unity. This allows the determination of a minimum value
for $b_\infty$. Setting an upper limit on it is a more
delicate affair and requires inspecting the calculated $\rho_i$ for
spurious oscillations induced by a too-high value of $b_\infty$.

The six local densities of states obtained after averaging over all
sites of a given coordination are shown in Fig.2.
The two curves correspond to the LDOS obtained for the 239 site approximant,
with eight nontrivial $b_n$ in the continued fraction, and for 1393 sites
 with 20 nontrivial $b_n$ and, clearly, more fine structure. This is to
be expected because replacing the exact $b_n$ by $b_\infty$ corresponds to
replacing the true quasicrystalline environment by some averaged effective
medium. Retaining 20 exact coefficients in the continued fraction yields the
exact quasicrystal upto a greater distance, hence to a finer energy
resolution. The tendency
is for A and B sites to have their greatest weight at the band edges, while
the E and F sites are most active in states at the band center. Not
shown are the individual LDOS for each particular site. These differ
from the averaged curves in
that sharp spikes appear superimposed on the overall form, in "random"
positions along the energy axis. We note also that that the density of
states thus found is in good accord with that determined by numerical
diagonalization.

The point to be stressed about these LDOS and their smoothed forms (with
increasingly wriggly shapes as one calculates further $b_n$ exactly in
the continued fraction) is not that there are only six different kinds
of sites in the quasicrystal (which naturally is quite false), but that
local environments are helpful in
understanding certain properties in a statistical sense. Thus, one may
expect to find that A sites have in general small amplitudes on
mid-band wavefunctions. In fact "localized" or "confined" states of zero
energy
can be constructed readily as for the Penrose tiling \cite{jap2}, and lead
to the enormous degeneracy found
at E=0. These are the wavefunctions that are particularly concentrated on
the F sites surrounding each of the A sites, leading to the peak that
one finds in the F site LDOS at E=0. Compared to the density of states
calculated by diagonalizing the Hamiltonian, the smoothed version
represented by our calculation is preferable on physical grounds for two
reasons: firstly, the presence of disorder in a real quasicrystal will
probably act to smoothen the jagged edges in the density of states of a
perfect quasicrystal. Finite temperature will also make it impossible to see
the selfsimilar structure at very small energy scales.
Thus it makes sense to
look at smoothed out functions. Second, the boundary conditions imposed when
diagonalizing the Hamiltonian always frustrate the tiling, and thus distort
the energy spectrum, leading for example to spurious peaks in the onsite
susceptibility (see below), if one is not careful.

Now we turn to the calculations of electron spin susceptibility in this
model, using the relation
\begin{equation}    \label{chi}
\chi_{ij}(E_F) ={1\over \pi}\int dE Im [G_{ij}(E) G_{ji}(E)]
sgn(E_F-E)
\end{equation}
where $N$ is the number of sites, and $i$ and $j$ site indices. This is the
standard formula for spin susceptibility written in real
space coordinates. $\chi_{ij}$ gives the spin polarization at site $i$
induced by a small magnetic field at site $j$. For the
onsite susceptibility $\chi_{ii}(E_F)$, we have used, first, Eq.\ref{ctfr}
of the Green's function, which yields a smoothed function
 similarly to the LDOS in Fig.2.
Second, we have calculated it directly by finding the eigenvectors of
the Hamiltonian, using mixed boundary conditions,
 for 41, 239 and 1393 sites, and using a discrete version of Eq.\ref{chi} .
The two methods yield
the same envelope curves for the local susceptibilities and are shown in
Figs.3 for each site type (averaged over sites of a given
type). The susceptibility is roughly flat for A and B
sites, while the F sites are
"more susceptible" for $E_F$ close to zero.

Next we examine the nonlocal susceptibilities $\chi_{ij}(E_F)$ as
a function of the distance of site j
 from a given site i. These oscillate, although not with a
defined
period as do the RKKY oscillations in a crystal. However the number of
zero crossings of the susceptibility increases with the Fermi energy,
reaching a maximum at $E_F=0$. Figs.4a) and b) show the oscillations for two
different values of $E_F$, over a range of distance that.
corresponds to half the lattice size of the
1393 site tiling. Also indicated is the $1/r_{ij}^2$ envelope of these
"quasi-RKKY-oscillations" -- which are an amusing demonstration of
 metallic character in the quasicrystal.

The effect of an onsite Coulomb repulsion U is now
considered.
We have treated this in the random phase approximation (RPA), and calculated
numerically the site susceptibilities given by the equation
\begin{equation}   \label{rpa}
\chi_{RPA} = \chi_0 (1 - U \chi_0)^{-1}
\end{equation}
$\chi_{RPA}$ being the RPA susceptibility matrix, and $\chi_0$ the
susceptibility matrix of Eq.\ref{chi}. This equation represents the
real space version of the Stoner equation \cite{mat} leading to itinerant
ferromagnetism in metals. A magnetic
instability is signalled when an eigenvalue of this matrix diverges. This
occurs when $U=\lambda_{max}^{-1}=U_c$ where we have determined
$U_c(E_F)$ numerically. Although it does not vary smoothly, $U_c$ is about
0.5 at the band center, increasing to values of the order of 1
 as the band edge is approached.
 The sites that are "most
susceptible" to going magnetic are found by studying the eigenvectors of
$\chi_0$ corresponding to $\lambda_{max}$. These are, speaking
statistically, almost exclusively
the F sites when $E_F$ is close to zero (about 25 \% of these sites), and
include all of the A and B sites when $E_F$ is near the band edges. A more
detailed
look at the distribution of susceptibilities will be given elsewhere.
For completeness, we may mention that Deng and Carlsson \cite{car} have
calculated
magnetic moments of Mn atoms in a cluster of AlMnSi, finding two types of
site carry no moment, while a third type of site does. This last case is
distinguished by the presence of a nearest neighbor Mn. This shows perhaps
the importance of including such pairwise interactions if one wishes to
study the distribution of moments (which is outside the scope of the present
paper).

Thus, numerical diagonalization and the continued fraction
technique yield similar results, showing the applicability of the latter
method to this quasiperiodic system, despite the singular features in the
energy spectrum. The smoothing features of the latter method are in fact
useful in helping identify overall features which could be obscured by the
irregular spikes of the exact solution ( which depends
on tiling size and
boundary conditions). There is seen to be a progression in the local density
of states and the susceptibility of individual sites, and A and B sites
behave quite differently from E and F sites, while the remaining two have
in-between character. Coulomb interactions enhance the
magnetic susceptibility of a given class of sites, depending on
Fermi energy, and could lead to the observed NMR observation of
nonhomogeneous values of the local moment.
 Finally, the mechanism that
is responsible for RKKY or Friedel oscillations in crystalline metals gives
rise, in the quasicrystal, to changes of sign of the susceptibility as a
function of distance, the frequency of which is determined by $E_F$. For
nearest neighbor sites, the "quasi-RKKY" interaction is initially positive,
but goes negative upon raising the Fermi level, staying negative all the way
until the upper edge of the band. For sites that are several bond
lengths apart, on the other hand, $\chi_{ij}(E_F)$ is a chaotically behaved
function of the Fermi energy.

 \noindent
{\em Acknowledgements --} I thank D. Spanjaard and M.-C. Desjonqu\`eres
for introducing me to the continued fraction representation, M.
Duneau for many useful discussions, and H. J. Schulz for valuable
comments and suggestions.


\begin{thebibliography}{10}
\bibitem{note}
A rather simple calculation gives one a first estimate of
the
band width: the infinite system of coupled equations contained
in the
tight-binding eigenvalue equation $H\vert \Psi_0\rangle = E_{max} \vert
\Psi_0\rangle$ ( $\Psi_0$ everywhere positive)
is truncated to a small set of
approximate equations
written for the six local environments, which can be solved for the energy
$E_{max}$ analytically. The value (necessarily approximate) obtained agrees
with the result of numerical diagonalization $\approx 4.2$.
\end{thebibliography}

Figure Captions \\
1. 239-site tiling with local environments labeled.
\\
2. Local density of states. Two curves are shown for each site type,
corresponding to taking 8 and 20 non-equal $b_n$ in Eq.\ref{ctfr} and taking
$b_\infty=4.48$.\\
 3. $\chi_{ii}$ versus $E_F$. The continuous curves show the
result using continued fractions, while the points are the
result of numerical diagonalization, for the 239 site tiling.
\\
4. Quasi-RKKY oscillations along a path of 19 sites (running index j),
taking i to be an A site on the 1393 site tiling. a) shows the slower
oscillations at the lower end of the energy spectrum and b) the rapid
oscillations near the band center. Dashed lines indicate $1/r^2$ decay.
The continuous lines are simply a guide for the eye.
\end{document}